\documentclass[aps,prb,twocolumn,groupedaddress,floatfix,showpacs]{revtex4-2}

\usepackage[T1]{fontenc}
\usepackage{graphicx}
\usepackage{amsmath}
\usepackage{color}
\begin{document}

\title{Entropy of timekeeping in a mechanical clock}

\author{David Ziemkiewicz}
\email{david.ziemkiewicz@utp.edu.pl}

 \affiliation{Institute of
Mathematics and Physics, UTP University of Science and Technology,
\\Aleje Prof. S. Kaliskiego 7, 85-789 Bydgoszcz, Poland.}

%\date{\today}

\begin{abstract}
The dynamics of an unique type of clock mechanism known as grasshopper escapement is investigated with the aim of evaluating its accuracy in a noisy environment. It is demonstrated that the clock's precision scales linearly with the rate of its entropy production, consistently with recently reported results regarding nanoscale and quantum clocks. Moreover, it is shown that the inevitable force variations present in the mechanism can be modelled with a Maxwell - Boltzmann statistic. Finally, the function of clock error is compared with Brownian motion and its fractal-like properties are discussed. The numerical results are confirmed with experimental data.
\end{abstract}
\maketitle

\section{Introduction}
Mechanical clock is a device that uses a driven oscillator (usually pendulum or balance wheel) for timekeeping. Its central part, the so-called escapement mechanism, provides energy to the pendulum at some fixed oscillation phase range, so that the pendulum becomes a self-excited oscillator \cite{Denny}. Various types of escapements have been analysed; some recent studies include anchor escapement \cite{Denny,Headrick}, gravity escapement \cite{Kesteven,Stoimenov}, verge and foliot \cite{Blumenthal,Danese}. Large overview of the basic physics of clock operation is presented in \cite{Rawlings}. To sustain the motion of the oscillator in the presence of dissipative forces such as air resistance, the escapement mechanism needs to continually input energy to the system. That energy is then dissipated into environment, raising its entropy and making the clock an irreversible system. In fact, the requirement of energy dissipation seems to be a fundamental feature of all clocks, including quartz oscillators, quantum periodic and non-periodic clocks \cite{Milburn}. To quantify the passage of time, one has to observe the evolution of some system toward higher entropy state. One question that immediately arises is how the amount of dissipated energy affects the precision of the clock. In the paper by P. Erker \emph{et al} \cite{Erker}, this issue has been extensively explored in a quantum system. It has been shown that there is a fundamental, linear relation between the accuracy and the rate of entropy increase. In \cite{Pearson}, this connection has been confirmed to exist in nanoscale, electromechanical clocks, which are semi-classical systems. In this paper, we attempt to establish some general relations between the rate of entropy increase and timekeeping accuracy, as applied to a mechanical clock. Naturally, such a classical, macroscopic system cannot approach the strict quantum bounds of accuracy; nevertheless, such an analysis may shed some new light on the dynamics and limitations of mechanical clocks. In particular, the grasshopper escapement, which is used in the world's most accurate mechanical clock \cite{McEvoy}, is analysed.

In 1622, John Harrison introduced a new type of mechanical clock mechanism, the so-called grasshopper escapement \cite{Harrison}. In many ways, its design seemed to be contradictory with established clockmaking principles and it was unclear how such a construction can attain higher precision than contemporary designs \cite{Aydlett}. Its most striking feature was a large pendulum amplitude; in contrast to the usual approach where the design strives to approximate an ideal mathematical pendulum, the Harrison's clock took advantage of the nonlinear effects emerging at large swing angles. One of the consequences of such a design is its relatively large power consumption - pendulum with a large oscillation amplitude has a greater average velocity and thus loses more energy in each cycle due to the air friction. While these losses can be avoided to some degree, removing them completely defeats the purpose of the clock - the pendulum has to achieve some steady state with finite amplitude. Harrison properly predicted that losses may not be detrimental to the accuracy, unknowingly stumbling upon a fundamental law that has been recognized almost 400 years later.

The timekeeping performance of grasshopper escapement is extensively studied in \cite{Millington1,Millington2,Hobden}. In our previous paper \cite{DZ_PRE}, it is shown that chaotic motion can emerge in such a system, introducing quasi-random disturbances to the pendulum's motion, even in absence of external influences. In this paper, the effects of such a random noise on the clock accuracy are discussed and the relations between noise level, clock power and accuracy are explored. The theoretical predictions and numerical calculations are are experimentally verified with a model, mechanical clock.

\section{Theory}
Consider a pendulum in a form of a point mass $m$, hanging on a weightless string with length $L$, that has been displaced from its resting position by an angle $\alpha$. The equation of its motion is
\begin{equation}\label{eq:mot0}
\ddot{\alpha}=\frac{M}{I}=\frac{-mgL\sin(\alpha)-\gamma\dot{\alpha}+M_E(\alpha)+M_N(t)}{I}
\end{equation}
where $M$ is the total moment of force and $I$ is the pendulum's moment of inertia, $\gamma$ is the damping coefficient, $M_E$ is the moment of force (torque) provided by the escapement mechanism and $M_N(t)$ is a noise term. For the grasshopper escapement, one can use an approximation \cite{DZ_PRE,Hoyng}
\begin{equation}\label{eq:mot1}
M_F(\alpha)=M_{F0}\mbox{sgn}(\alpha-\alpha_r\mbox{sgn}(\dot{\alpha}))
\end{equation}
where $M_{F0}$ is a constant and $\alpha_r$ is a geometrical parameter - the pendulum angle at which the driving torque switches sign. It should be pointed out that the forcing term $M_F$ is not explicitly time-dependent, but rather tied to the pendulums position and thus it has the same frequency as the pendulum itself. This makes the system a self-excited oscillator. In particular, its an example of van der Pol's oscillator \cite{Hobden}. Such a system exhibits a characteristic local minimum of oscillation period, making it insensitive to small changes of driving term (in case of clocks - torque) \cite{Akcasu,Millington1,DZ_PRE}.

The noise term $M_N(t)$ is a phenomenological quantity encompassing multiple processes. In \cite{DZ_PRE} it has been shown that chaotic motion in the system leads to small, random disturbances of period; the plot of total error, e.g. the sum of errors of individual periods, exhibits structure resembling a random walk. One of the possible forms of $M_N$ that lead to such a dynamic is white Gaussian noise, which is also a close approximation of thermal noise in electrical systems \cite{Garnier}. As it will be shown in the later part of the manuscript, in the case of mechanical clock a better model can be proposed.

The Eq. (\ref{eq:mot0}) with the forcing term (\ref{eq:mot1}) is solved numerically by integrating the equation of motion with a fixed, discrete time step \hbox{$\Delta t=20$ $\mu s$.} The chosen value is a compromise between simulation time and accuracy. The noise is a random variable added at every time step. Due to the fact that the pendulum period is on the order of one second, the noise changes happen at a timescale that is many orders of magnitude faster. On the other hand, long-term effects originating from environmental changes of temperature, pressure etc. \cite{Nelson} are not explored here. 

\begin{figure}[ht!]
\includegraphics[width=.95\linewidth]{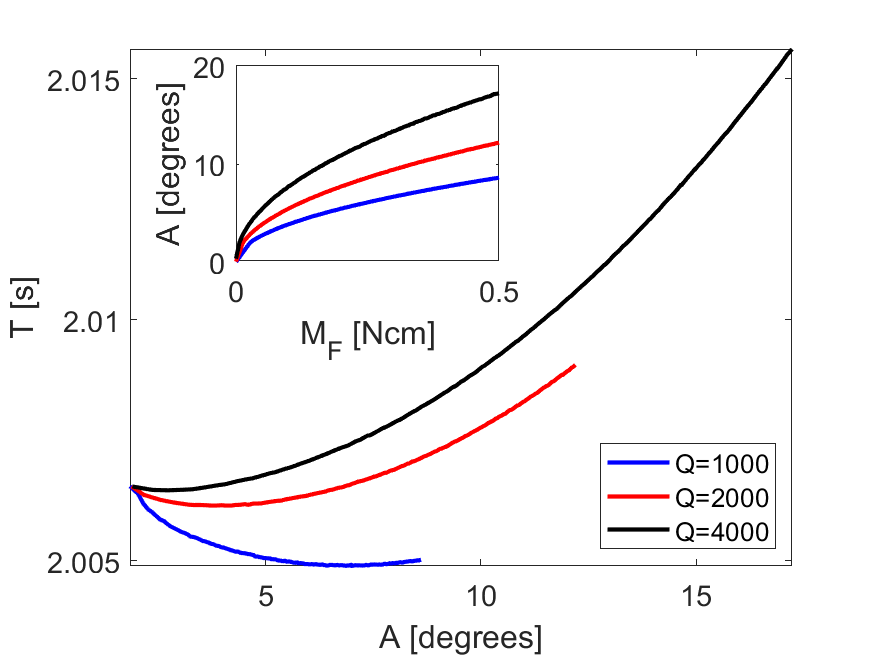}
\caption{The pendulum period as a function of driving torque $M_F$, calculated for 3 values of Q factor. Inset: amplitude as a function of driving torque.}\label{fig_1b}
\end{figure}

A characteristic feature of grasshopper escapement is the existence of the local minimum of the function $T(A)$, which is a result of interplay between the effect of escapement and the increase of physical pendulum period with its amplitude. An accurate estimation of this effect is presented in \cite{Millington1}. Near this minimum, $\partial T/\partial A \approx 0$ and so the system is insensitive to small changes of amplitude. The location of this minimum depends on the Q factor of the pendulum and the geometry of the escapement, expressed by the parameter $\alpha_r$. The typical function $T(A)$ obtained in numerical simulation is shown on the Fig. \ref{fig_1b}.
Notably, for the lower $Q$ factor the local minimum is wider, which indicates that such a system is more tolerant to small changes of amplitude (and thus torque). It is beneficial to have a more lossy system. However, this result applies to the long-term effects; the period values shown on the Fig. \ref{fig_1b} are obtained in a steady state, when the power delivered to the pendulum is equal to the power lost. In such a case, the system's trajectory is a so-called limit cycle \cite{Libre}. However, momentary forces due to the noise move the pendulum away from its limit cycle and thus affect the period. In particular, lets consider a short impulse (shorter than the period) that delivers some energy $\Delta E$ to he pendulum, disturbing it from its steady-state operation. The increase of kinetic energy by $\Delta E$ causes a change of mean velocity
\begin{equation}\label{eq:dedv}
\Delta V \sim \frac{\Delta E}{V} \sim  \frac{\Delta E}{\sqrt{E}}.
\end{equation}
Therefore, a general rule, regardless of the particular type of clock mechanism used, is that for a given power of noise, a pendulum with larger mean velocity, and thus larger energy, experiences smaller change of mean velocity and, correspondingly, period. To reduce the errors caused by random noise, one should aim for the largest practical pendulum energy. Moreover, as shown in \cite{DZ_PRE}, in systems with lower Q factor the effects of any disturbance die out faster, so that the pendulum returns to steady-state operation in a shorter time, producing smaller total timekeeping error (sum of period errors). The brilliance of grasshopper escapement lies in the fact that a pendulum with low Q factor and large amplitude is exactly the right choice to reach the local minimum $\partial T/\partial A$, so that the clock becomes insensitive to both long-term and short-term disturbances.

\section{Experimental results}
\begin{figure}[ht!]
a)\includegraphics[width=.55\linewidth]{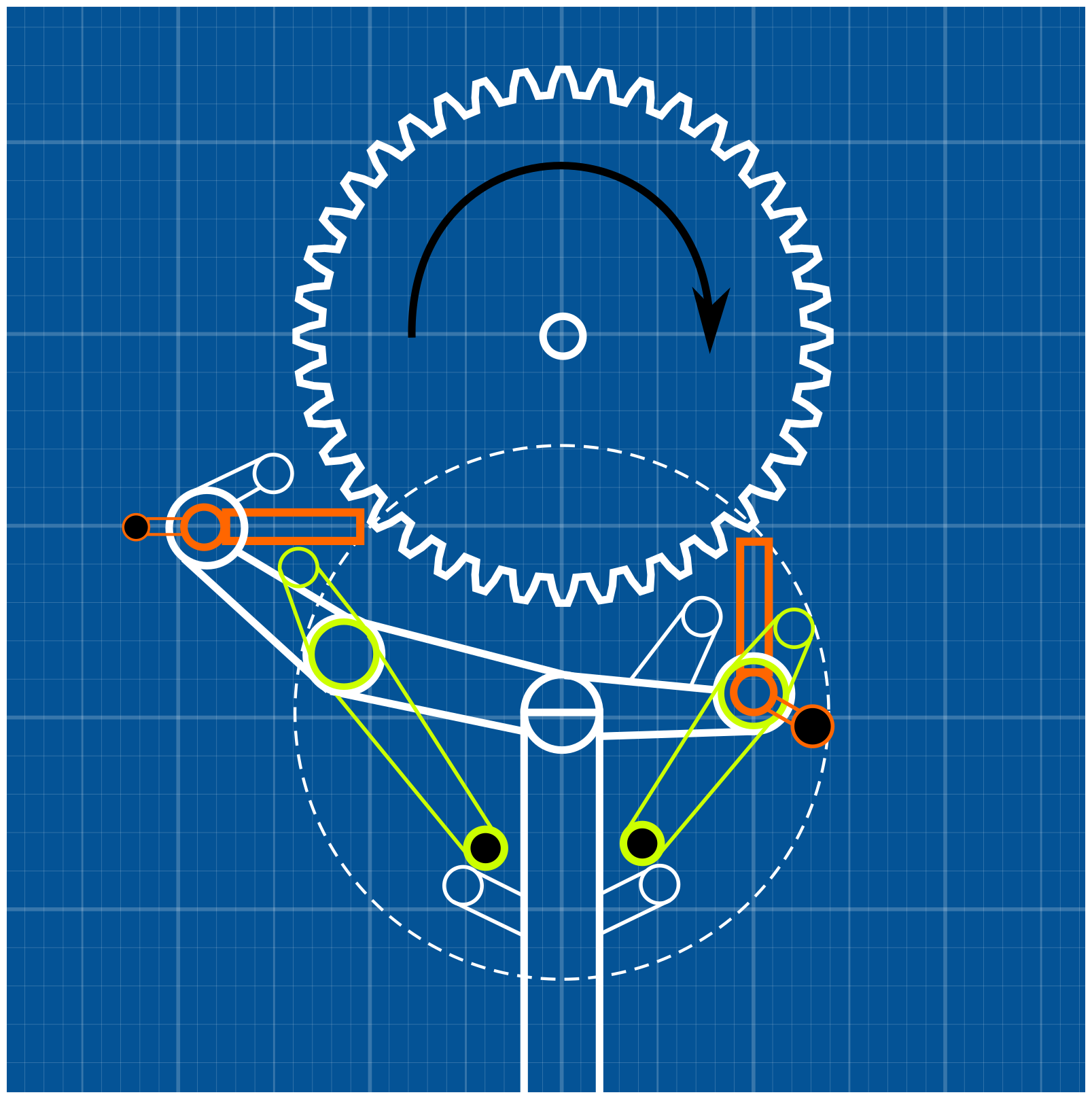}
b)\includegraphics[width=.35\linewidth]{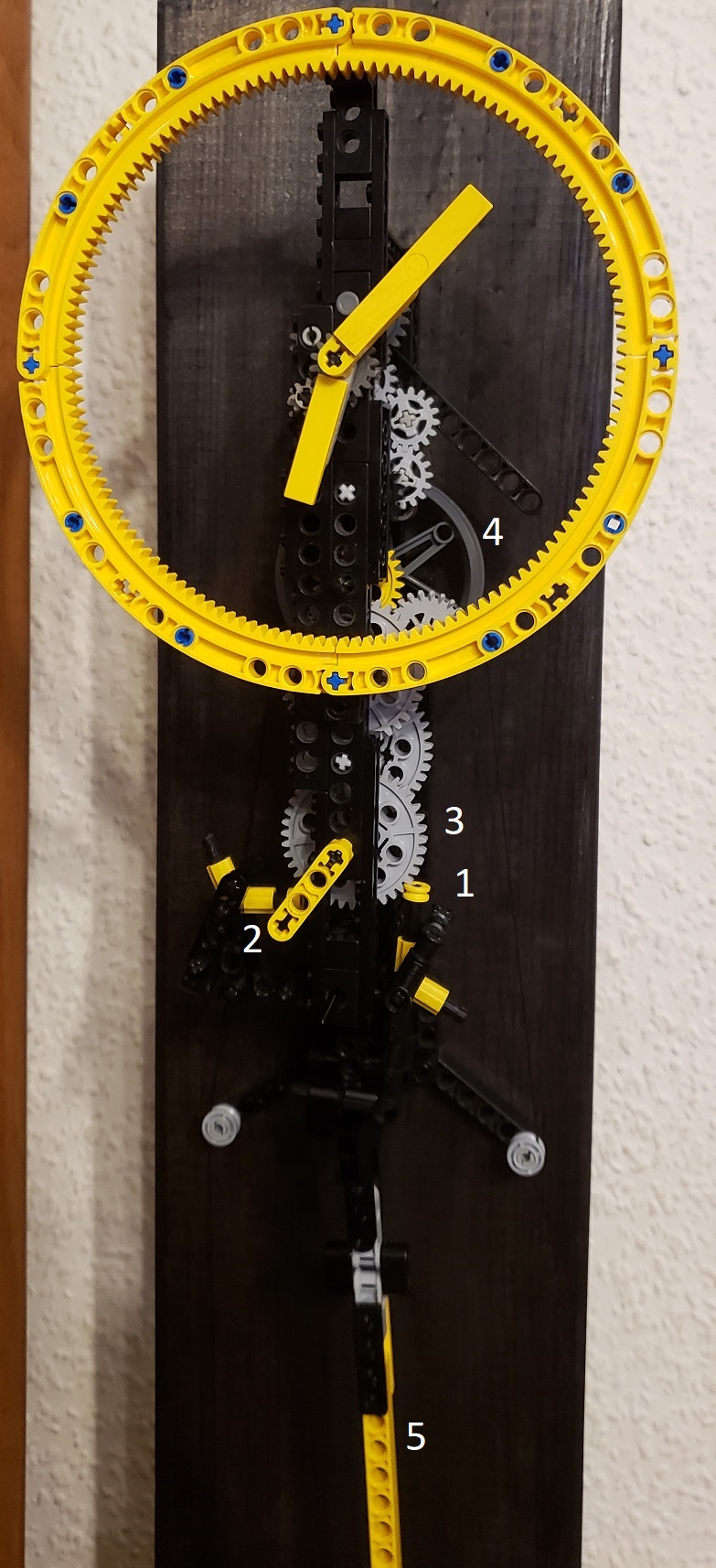}
\caption{a) Schematic of the grasshopper escapement: escape wheel (top) engages the pallets (orange); the pallet resting positions (green) are moving, allowing for recoil. b) Realization of the escapement with LEGO bricks; pallets 1, 2 mesh with the escape wheel 3, providing an interface that transfers power from the spool 4 to the pendulum 5.}\label{fig:esc}
\end{figure}

\begin{figure}[ht!]
a)\includegraphics[width=.9\linewidth]{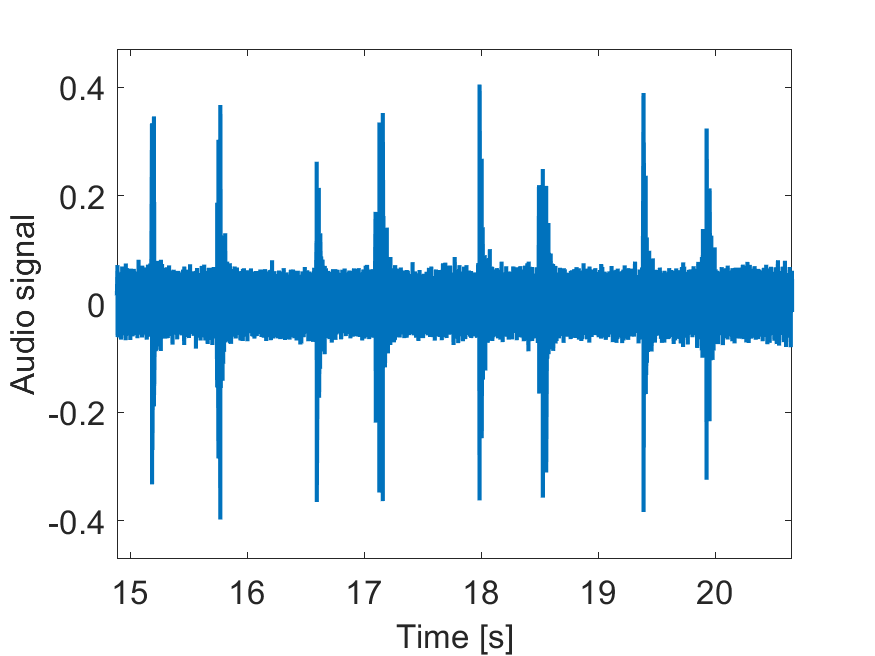}
b)\includegraphics[width=.9\linewidth]{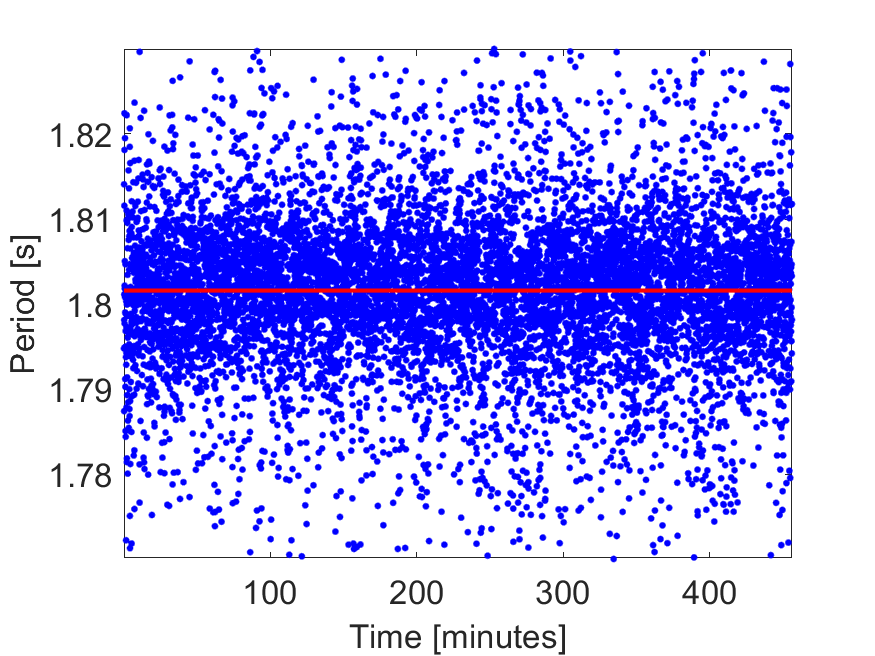}
\caption{a) Fragment of the typical sound recording obtained in the experiment. b) The calculated period values.}\label{fig_2}
\end{figure}

To verify the correctness of numerical simulation, a model clock fitted with grasshopper escapement has been constructed and its speed was measured. For ease of quick prototyping, LEGO bricks have been used in construction. While this choice may seem surprising at first, the high accuracy standards to which the individual elements are made allows for the creation of highly precise instruments, such as Watt balance capable of measuring the Planck constant with less than 1$\%$ error \cite{Watt}.
The central part of the clock is the escapement mechanism, which is shown on the Fig. \ref{fig:esc} a). In contrast to the usual implementation of such a mechanism \cite{Aydlett}, the pallets (marked with orange color) that engage the escape wheel are located below it and are connected directly with the pendulum, which simplifies construction without impacting the escapement's general characteristics. The interface points between left/right pallet and escape wheel are located at the same distance from pendulum pivot (center of dashed circle), ensuring that the moment of force acting on the pendulum is the same whether it moves to the left or to the right. Overall, the moment of force is not constant and varies slightly with pendulum angle, but this is not an important factor in system dynamics \cite{DZ_PRE}. The Fig. \ref{fig:esc} b) shows the LEGO implementation of the mechanism. The pendulum (marked with number 5) has a period of 1.8 seconds, mass of 0.3 kg and an estimated $Q \approx 1500$. The clock is powered by a weight on a rope, unwinding from a spool (4) and providing torque. The driving weight of $m=0.2$~kg descends at a rate of 20 cm per hour, providing approximately 110~$\mu$W of power. For more details, see \cite{filmy}.
The period of oscillation is measured by recording the ticking sound of the clock. In every period, two audible ticks can be heard, corresponding to the moment when left/right pallet disengages the escape wheel and falls down to its resting position. An example recording is shown on the Fig. \ref{fig_2} a). 
The sampling rate of the audio signal is 44 kHz, providing a theoretical upper limit of the time resolution $\delta t \approx 23 \mu$s. Due to the fact that the escapement mechanism is not fully symmetric, the peaks corresponding to left/right pallet are not equal. Furthermore, one can see some period to period variation of the peak height. This indicates that the force acting on the pallets is not constant in time, but exhibits some random variation. The period is calculated by separating the peaks from the background using fixed threshold and calculating the time between peaks. The threshold is fitted to ensure a good compromise between the fraction of captured peaks and false signals due to the background noise. Any outliers (periods that differ from the mean by more than 5$\%$) that may result from the said noise, are discarded. The calculated periods are shown on the Fig. \ref{fig_2} b). The nominal period of the pendulum is 1.8 s. The average value (red line) is slightly larger, indicating that small correction of the pendulum length is necessary. The measured periods seems to follow a normal distribution with a standard deviation of $\sigma \sim 0.009$ s.

\begin{figure}[ht!]
\includegraphics[width=.95\linewidth]{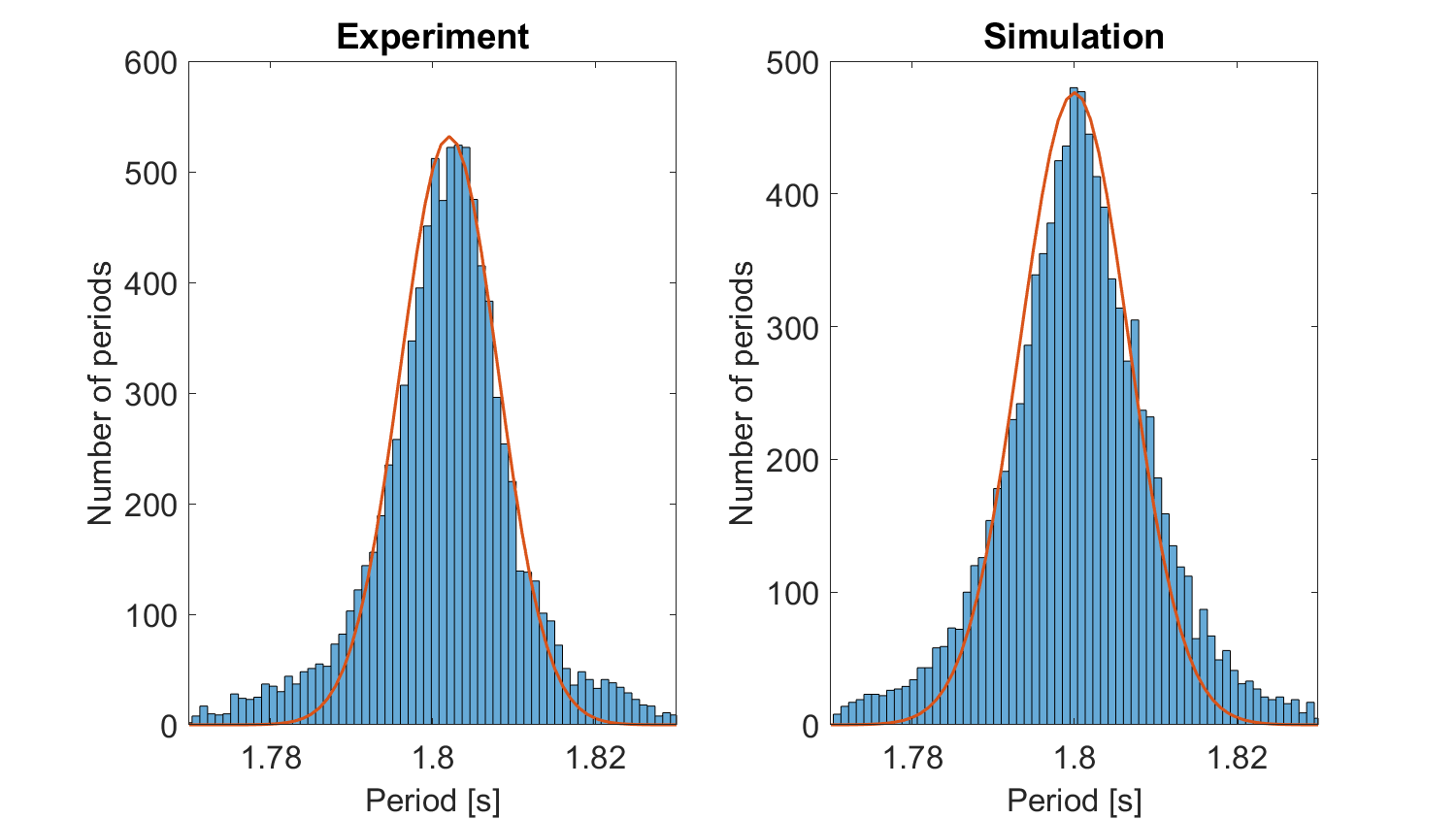}
\caption{Comparison of the period distribution obtained in experiment (left) and simulation (right).}\label{fig_3}
\end{figure}

In order to verify the hypothesis that the period variance can be modelled with a white noise, the experimental data has been compared with numerical simulation results. The comparison of period distributions is shown on the Fig. \ref{fig_3}. 
On the left panel, one can see the period histogram (bars) and a fitted Gaussian (red line). There is a fairly good match in the region near the average value, but in the data there are more values on the sides of distribution. In other words, periods that differ considerably from the mean one are slightly more probable than in normal distribution. There are several factors contributing to such a result. The pendulum is a system that is characterized by some memory of past events; any given period is affected not only by disturbances happening in that period, but also to ones that occurred in past periods. Thus, the underlying condition that periods are statistically independent is not exactly met. Moreover, the noise itself, representing variation of driving torque, originates from interaction between many mechanical elements of the clock and may contain many short-term and long-term contributions. Naturally, direct modelling of the contact friction between all clock parts (gears, axles) that form the interface between the source of power and pendulum is challenging. One of the approaches, proposed in \cite{Moon} and adapted in our previous paper \cite{DZ_PRE} is to model the mechanism as a harmonic oscillator with some resonant frequency, which is coupled to the pendulum. Such a double-oscillator system exhibits chaotic motion \cite{Levien}, resulting in quasi-random driving torque. Chaotic motion can also emerge in some single-oscillator systems that are periodically driven \cite{Kobes} or parametrically damped \cite{Smith}. All these features are present in the dynamics of a mechanical clock.

Potential alternative approach, which will be explored here, is based on the fact that the clock mechanism consists of a large number of interacting parts that exchange energy. In particular, lets assume that the system contains gears that are freely moving between collisions (e.g. the majority of friction comes from meshing teeth, not from axle bearings). In such a case, one can propose that the noise function follows a Maxwell-Boltzmann distribution, which is derived under the same assumption that we have a large number of particles that interact only during collisions. In particular, one can use a distribution of velocity in single axis
\begin{equation}\label{eq:MB}
f(V)=\sqrt{\frac{m}{2\pi kT}}\exp\left(\frac{-mV^2}{2kT}\right)
\end{equation}  
where $m$ is particle mass, $k$ is the Boltzmann constant and $T$ is temperature. In case of nano-resonator operating in dilute gas, the above relation can be used to model the force acting on the resonator caused by collisions with gas molecules; only single axis is used because it is assumed that the period is measured by observing the resonator motion in one axis; this is the case for mass on a spring, membrane and pendulum with low amplitude. Interestingly, in a lot of cases the Maxwell-Boltzmann statistic remains valid in dense fluids and even in solids \cite{Mohazzabi}. Moreover, the statistic remains valid with a good approximation even in cases where the number of bodies involved is relatively small, as demonstrated in \cite{Boozer}. This is the base of the assumption that such a distribution may be useful for modelling of mechanical clock. It should be stressed that in macroscopic system, the individual disturbances are also macroscopic and do not come from air molecules; thus the mass $m$ and temperature $T$ in Eq. (\ref{eq:MB}) are abstract fitting parameters, not properties of any physical body. For a numerical justification of the Eq. (\ref{eq:MB}), see Appendix A.  

The results of numerical simulation obtained with the above mentioned model are shown on the Fig. \ref{fig_3} right panel. The period histogram shows an excellent match with the experimental data on the left panel. Again, a fitted normal distribution underestimates the values far from the mean. The noise term is modelled with Eq. (\ref{eq:MB}); at every computation step, an uniformly distributed random value $V \in (0,10)$ is chosen and the noise term is given by $M_N=M_{N0} Vf(V)$, where $M_{N0}=100$ Ncm is the noise amplitude and the distribution parameters are $m=1$, $kT=0.9$. Due to the fact that the numerical time step $\Delta t=20$ $\mu$s is very short, individual noise values have a little impact on the pendulum motion despite the seemingly large value of $M_{N0}$.

\begin{figure}[ht!]
\includegraphics[width=.95\linewidth]{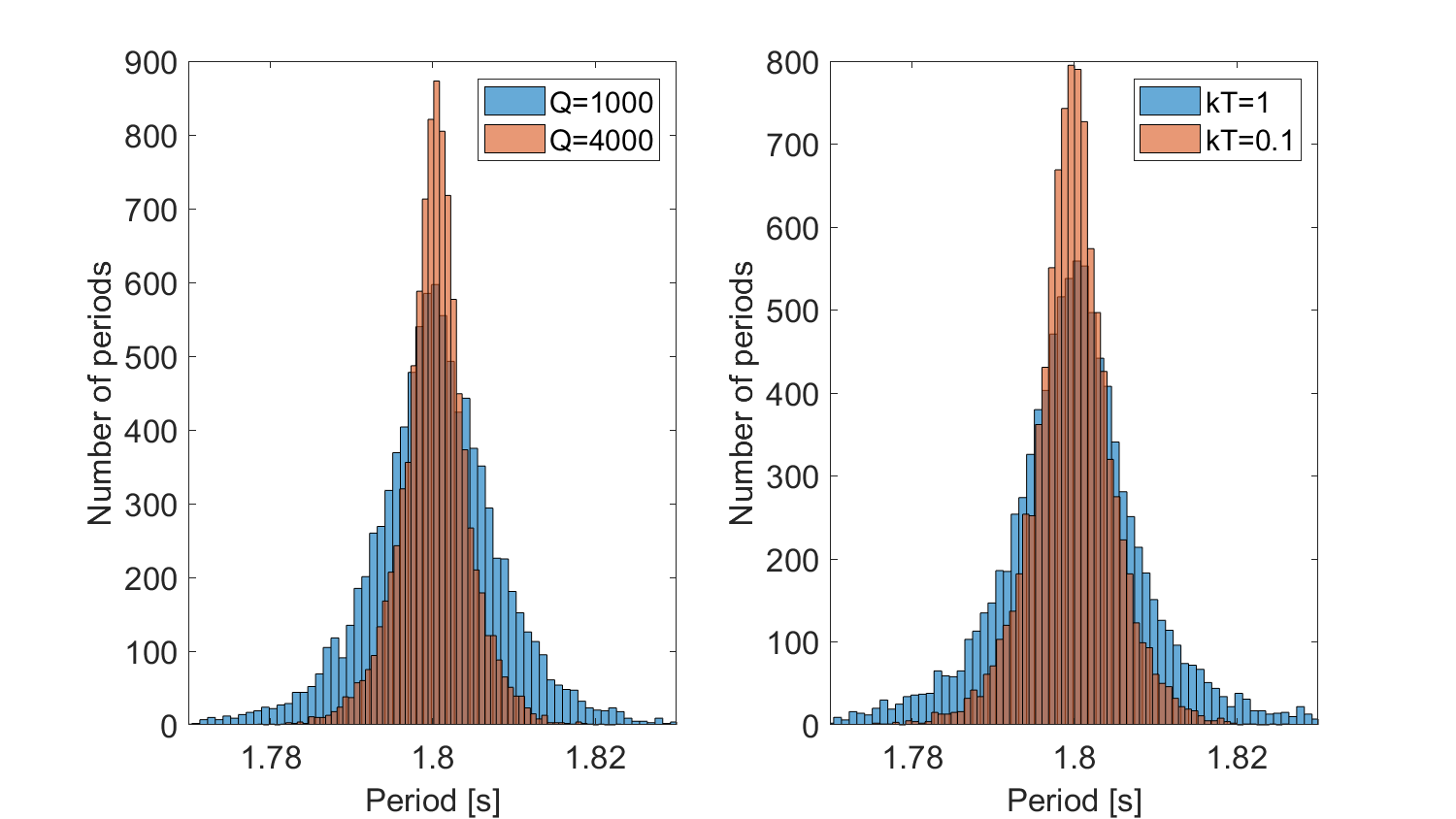}
\caption{Period histograms obtained for various values of pendulum Q factor (left panel) and temperature parameter (right panel).}\label{fig_3b}
\end{figure}

By changing the simulation parameters, one can make some general observations. On the Fig. \ref{fig_3b} the calculation results for selected values of pendulum $Q$ factor and temperature parameter $kT$ are shown. As expected, both decrease of Q factor and increase of temperature result in a wider spectrum. The results obtained in the limit of high Q factor and low temperature are closer to a normal distribution.

\begin{figure}[ht!]
\includegraphics[width=.95\linewidth]{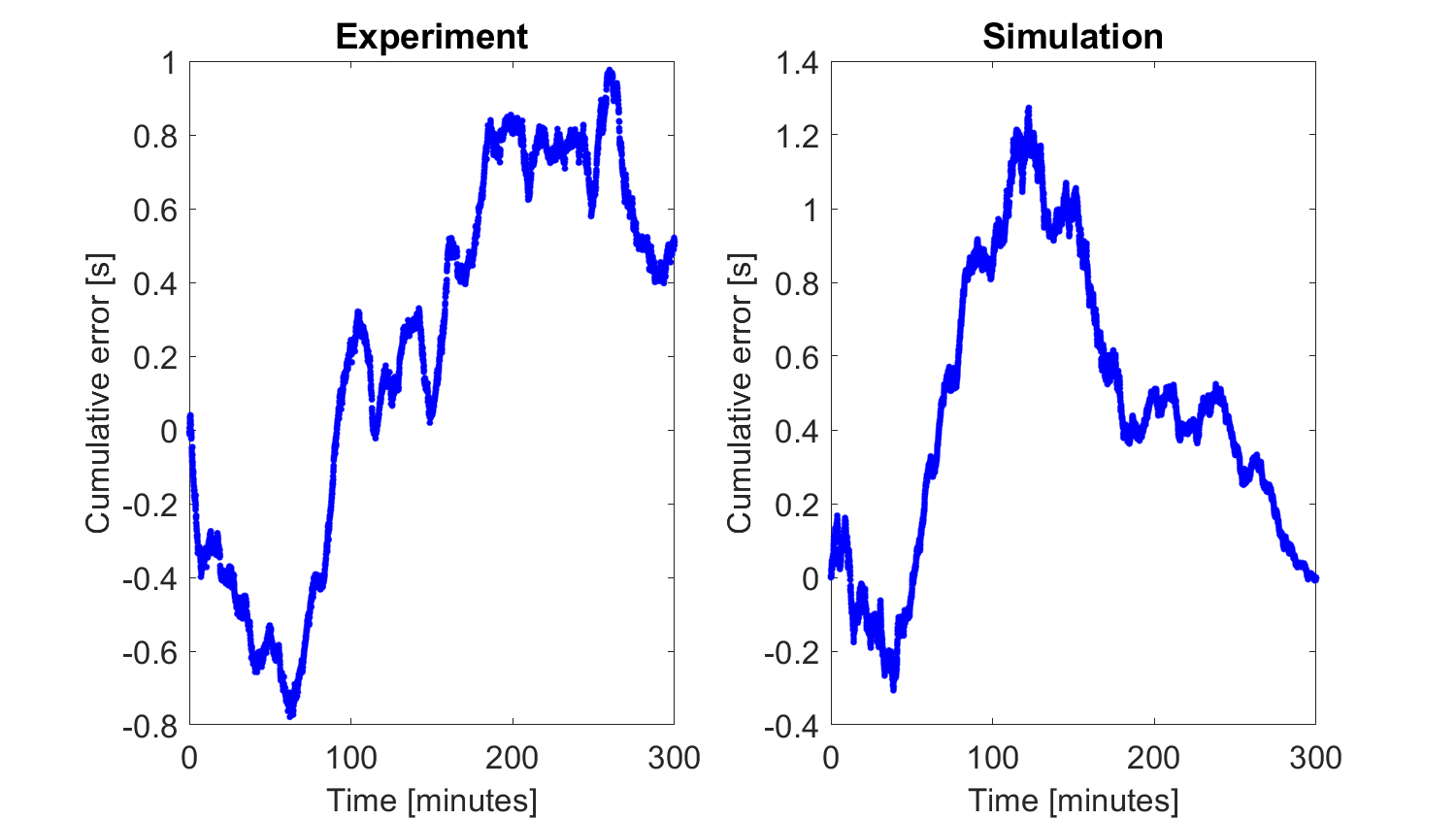}
\caption{Comparison of the total error obtained in experiment (left) and simulation (right).}\label{fig_4}
\end{figure}

On the Fig. \ref{fig_4}, the total clock error (the difference between clock time and reference time) is shown. Again, there is a close correspondence between experimental and numerical results. The plot resembles a random walk, which is a typical result in such measurements \cite{Hoyng}. The shape of the plot has a fractal-like structure that is self-similar; this feature will be discussed in detail later.
Overall, the results indicate that the assumed numerical model is correct and can be used to predict the dynamics of the clock.

\section{Energy dissipation and accuracy}
\begin{figure}[ht!]
a)\includegraphics[width=.9\linewidth]{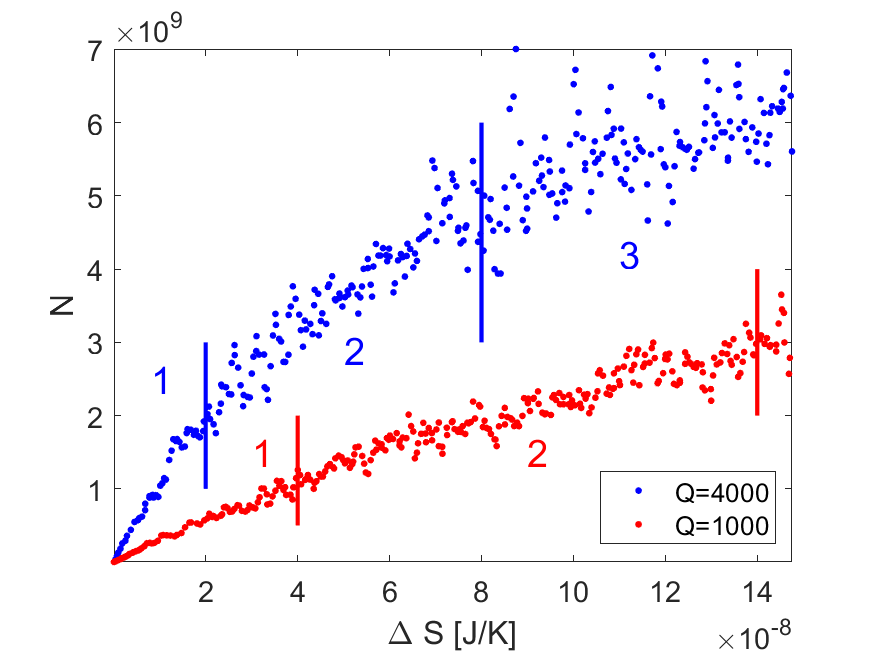}
\caption{The parameter $N$ as a function of entropy increase per period $\Delta S$, calculated for two values of $Q$.}\label{fig_5}
\end{figure}

\begin{figure}[ht!]
\includegraphics[width=.9\linewidth]{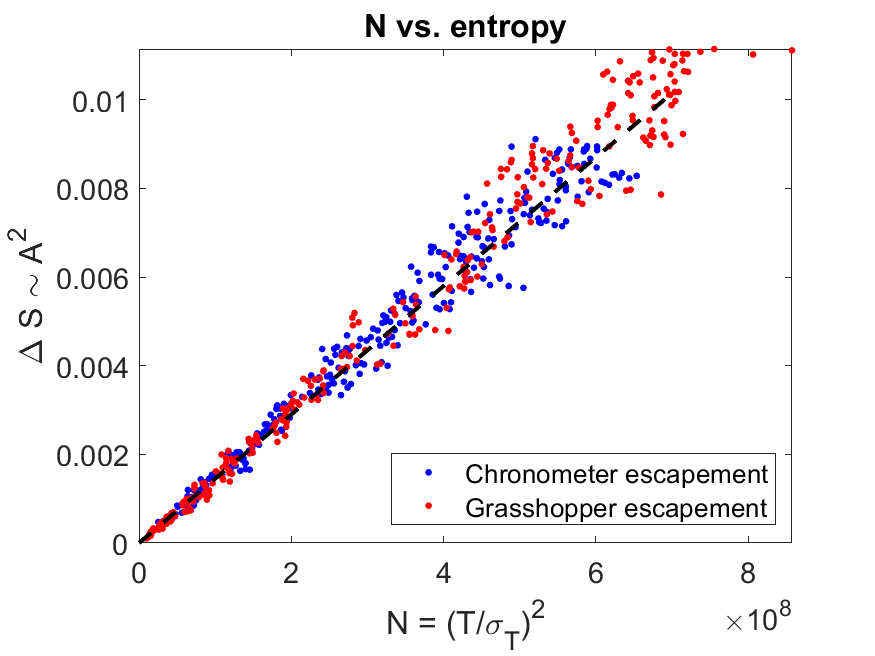}
\caption{The relation between parameter $N$ and entropy increase per period $\Delta S$, calculated for two types of clock mechanisms.}\label{fig_6}
\end{figure}

For a pendulum with a given $Q$ factor, the energy dissipated in a single period is given by
\begin{equation}
\Delta E = \frac{E}{Q},
\end{equation}
where $E$ is the total energy of the pendulum. For the amplitude $A$, $E=\frac{1}{2}I\omega^2A^2$. The energy is lost due to the pendulum suspension friction and air friction. In both cases, it ends up as the heat dissipated in the environment. The change of entropy in a single period is
\begin{equation}
\Delta S = \frac{I\omega^2A^2}{2QT}
\end{equation}
where $T$ is the temperature of the environment. One can reduce losses either by increasing the Q factor or reducing the amplitude and frequency. In realistic implementations, there is an upper bound to the Q factor depending on the quality of the suspension mechanism and air resistance of the pendulum. The frequency is tied with the length of the pendulum. For a concrete example, lets consider a pendulum with length $L=1$ m, mass $m=1$ kg, driven by grasshopper escapement with torque $M_{F0} \in (0,1)$ Ncm. The white noise has an amplitude of $M_N=0.1$ Ncm and the system operates at room temperature $T=300$ K.  
To qualify the accuracy of the clock, one can introduce a parameter $N$ \cite{Erker}
\begin{equation}
N = \left(\frac{\bar{T}}{\sigma_T}\right)^2,
\end{equation} 
where $\bar{T}$ is the mean period and $\sigma_T$ is the standard deviation of the period. The value of $N$ is the number of periods after which the standard deviation of the clock time (e.g. the sum of periods) is equal to a single period\cite{Erker}. The value of $N$ as a function of entropy increase per period, calculated for a range of torques and two Q factors, is shown on the Fig. \ref{fig_5}. 
In a steady state, the energy dissipated per period is equal to the work done by the escapement mechanism. Thus, the torque is connected with $\Delta S$ by
\begin{equation}
\Delta S = \frac{2}{T}\int\limits_{-A}^{A}M_F(\alpha)d\alpha.
\end{equation}
As the torque $M_F$ increases, both the amplitude $A$ and the work done in a single period are also increasing. In the Fig. \ref{fig_5}, for the $Q=4000$ (blue curve), one can notice three distinct regions marked 1, 2, 3 and separated by vertical lines. In the region 1, the power provided by the escapement is insufficient to sustain the motion of the pendulum. The steady operation happens in region 2. One can see that the relation between $N$ and $\Delta S$ is linear, which is consistent with results in \cite{Erker} and \cite{Pearson}. By further increasing the driving torque, one reaches the saturation region 3. In this regime of operation, the accuracy $N$ is still increasing with $\Delta S$, but at a slower rate. The dynamic becomes more chaotic, with larger variation of obtained $N$ values. The results calculated for smaller Q factor (red curve) show similar structure. However, due to the fact that for smaller $Q$, at any given amplitude $A$ more power is dissipated, the boundaries of region 2 and 3 are shifted towards larger $\Delta S$. Initially, results for $Q=1000$ are characterized by roughly 4 times smaller accuracy than $Q=4000$. However, due to the longer linear region 2, the peak accuracy before saturation happens is less than 2 times lower for the more lossy pendulum.
It should be mentioned that the local minimum $\partial T/\partial A$ is located near the beginning of the area 2 in the Fig. \ref{fig_5} and does not correspond to any characteristic point of the plot. One can set the operating point to an amplitude larger than the position of this minimum, resulting in greater $\Delta S$ and smaller sensitivity to noise, at the cost of larger impact of long-term effect due to $\partial T/\partial A>0$. In conclusion, the optimal choice of the amplitude (and so the driving torque) depends on the ratio of short-term to long-term disturbances present in the system. 
To further confirm that the characteristic shape of the $T(A)$ function has no direct impact on the noise sensitivity of grasshopper escapement, one can make a comparison with other type of mechanism such as chronometer escapement. In the chronometer escapement, the driving torque function $M_F(\alpha)$ has a form of a single, short impulse occurring at $\alpha \approx 0$, which is the point where the effect of such impulse on the period is smallest \cite{Rawlings,Hoyng}. The comparison results are shown on the Fig. \ref{fig_6}. One can see that both types of mechanism exhibit exactly the same, linear relation between $N$ and $\Delta S$ when matched with the same pendulum. This indicates that the obtained result is not a specific feature of the mechanism in question, but a more general tendency; the advantage of grasshopper escapement is in the possibility of using large amplitude, and thus large $\Delta S$, with no disadvantage to long-term effect sensitivity. 

\section{Fractal structure of the error function}
\begin{figure}[ht!]
\includegraphics[width=.9\linewidth]{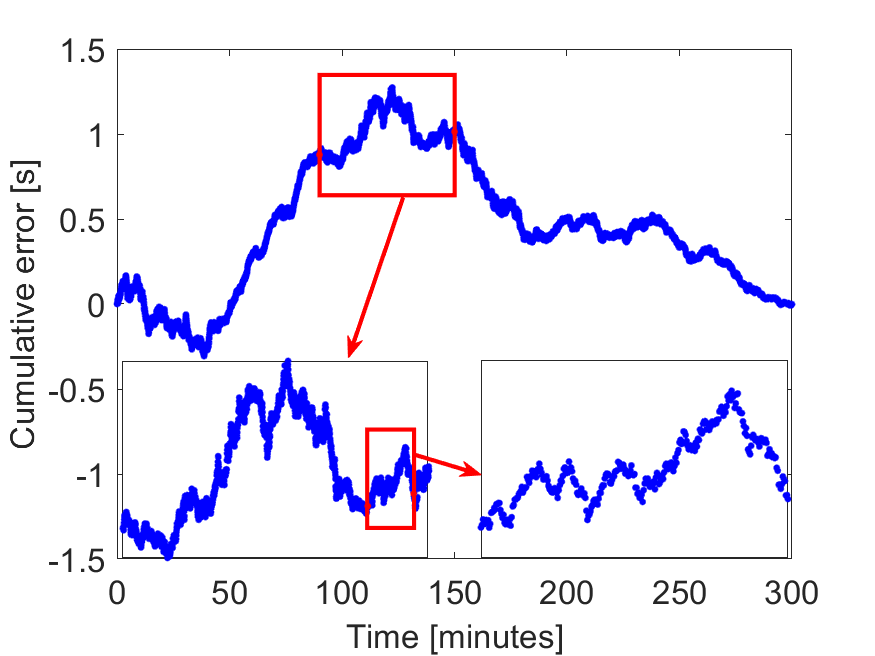}
\caption{The function $E_T(t)$ calculated from experimental data.}\label{fig_7}
\end{figure}

Lets assume that a perfect reference clock with period $T_R$ is used to evaluate the accuracy of the observed, noisy clock. The total error of the clock $E_T$ at a time $t$ is the sum of the errors 
that occurred in every period up to $t$. Thus, it can be defined as
\begin{equation}\label{eq:ET}
E_T = \sum\limits_{i=1}^{t/T_R}(T_i-T_R).
\end{equation}
As mentioned before, the function $E_T(t)$ appears to have a fractal-like structure. Its self-similarity is demonstrated on the Fig. \ref{fig_7}; by magnifying a small part of the plot, one obtains a fragment that is very similar to the whole curve.
On the lowest level, one has individual periods that are either longer or shorter than the reference period. Every disturbance produces a jump in the period value. In the case where there is no correlation between jumps, one obtains a random walk, which in the limit of a large number of small jumps approximates Brownian motion \cite{Knight}. However, as observed earlier, the individual periods cannot be treated as completely independent and thus the function $E_T(t)$ is only an approximation of trajectory of Brownian motion. To resolve the small differences between the clock error and true random walk, one can use the concept of fractional dimension $D$. The Brownian motion is characterized by the value of $D=1.5$ \cite{Falconer}.

\begin{figure}[ht!]
\includegraphics[width=.9\linewidth]{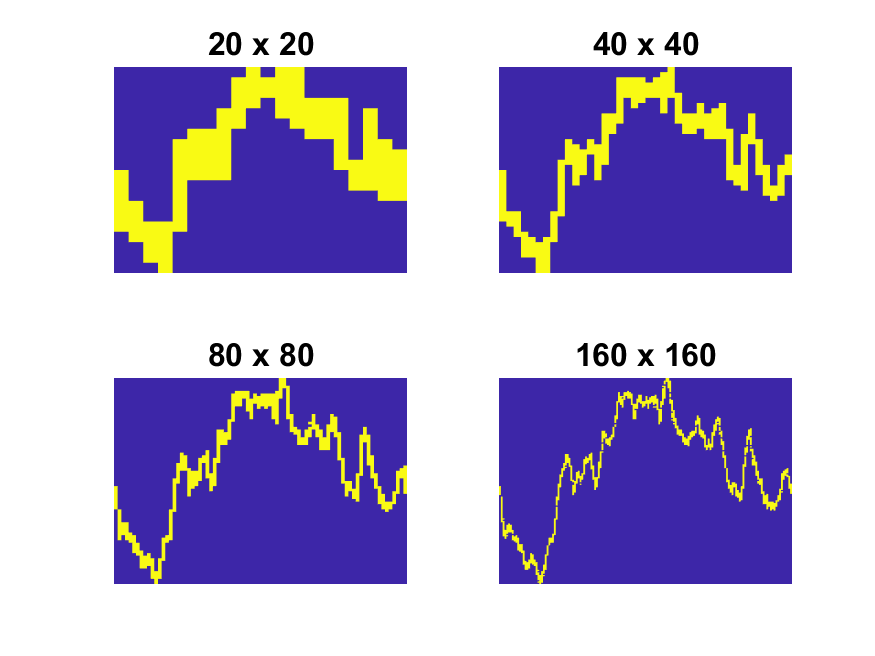}
\caption{Four iteration steps of the box-counting procedure. The curve is put in a space of a given size (text over the images) and the number of boxes containing the curve (yellow rectangles) is counted.}\label{fig_box}
\end{figure}

To calculate the dimension of the plot of the function $E_T(t)$, one can use the so-called box-counting method \cite{Schroeder}. Let's consider a measured or simulated set of $n$ periods, with values $T_i$,~$i=1..n$. Every period starts at some time $t_i$. By using the Eq. (\ref{eq:ET}), one can construct a series of values of total error $E_{Ti}$. The pairs of values $(E_{Ti},t_i)$ are interpreted as points on a two-dimensional plane. To perform the box-counting procedure, the plane containing the curve is divided into $N$ squares (boxes). Then, one counts the number $M$ of boxes which are non-empty, e. g. contain one or more points. The dimension is defined as
\begin{equation}\label{eq:dim}
D = D_0\frac{\partial \log M}{\partial \log M},
\end{equation}
where $D_0$ is the number of dimensions of the space that contains the curve (in our case, $D_0=2$). For example, if the size of the boxes is reduced by a factor of 2, then the number of boxes covering a 2-dimensional plane increases by a factor of $2^2$, while the number of boxes containing a one-dimensional line increases by a factor of $2^1$. The process is illustrated on the Fig. \ref{fig_box}. The counting is done iteratively; as the box size becomes smaller, the number of boxes containing the curve (marked by yellow rectangles) increases. The rate of this increase is proportional to the dimension. Specifically, one needs to calculate the slope of the function $\log M(\log N)$, as indicated by Eq. (\ref{eq:dim}).

\begin{figure}[ht!]
\includegraphics[width=.9\linewidth]{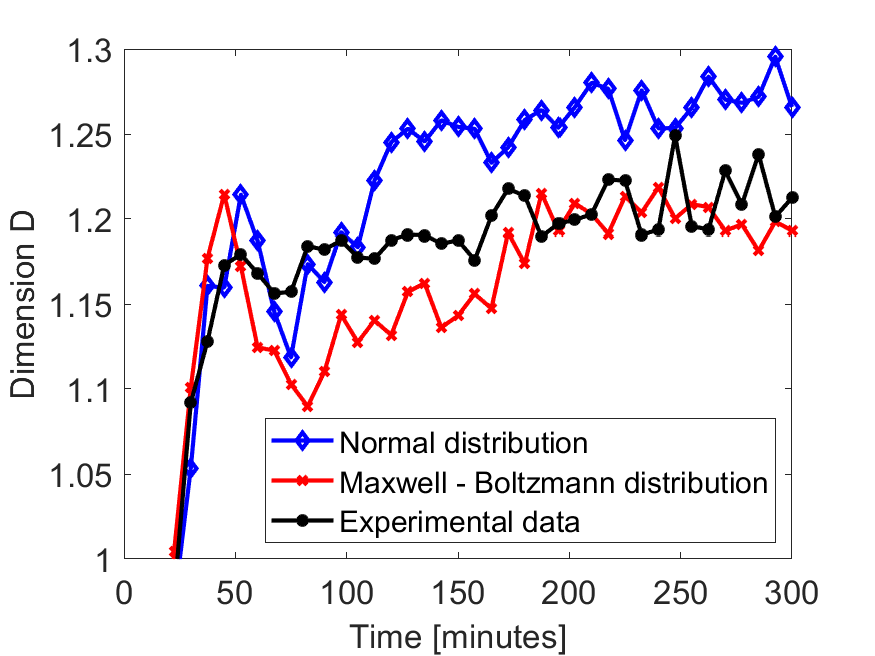}
\caption{The relation between the length of measurement/simulation and the dimension of the curve $E_T(t)$.}\label{fig_dimt}
\end{figure}

One important difference between strict mathematical fractals and fractal-like measured data is the fact that physical objects are not infinitely self-similar; there is some lowest scale, which in our case is a single period. A two-element data set containing two consecutive periods produces a plot that is a straight line connecting the points, which is one-dimensional. As the number of points increases, the curve becomes more complicated, containing more details on various time scales. Thus, in general one can expect that the dimension increases with the length of the data. This is the case in the measured and calculated results, as shown on the Fig. \ref{fig_dimt}. 

Both experimental and simulation results start from $D=1$ and converge upon some fixed value in the long time limit. Notably, the simulation results where the noise has been modelled with Maxwell-Boltzmann distribution provides a better fit to the experimental data. Specifically, both dimensions converge upon a value of $D \approx 1.2$, while normal distribution noise results in $D \sim 1.27$. The obtained dimension is significantly lower than the theoretical value of $D=1.5$ for the Brownian motion. However, it matches the results for the so-called fractional Brownian motion. Such a system is characterized by a so-called Hurst exponent, which is a measure of the long-time memory of the system. Specifically, the obtained data is characterized by a Hurst exponent $H=0.7$. The value of $H>0.5$ indicates that there is a positive correlation between consecutive periods and the system has a memory of past states \cite{Mandelbrot,Roman}. The process is persistent, e.g. the chance to obtain two consecutive periods that are both shorter or longer is increased \cite{Sacrini}. 

A question arises whether the noise description based on fractional Brownian motion is compatible with the approaches that treat the system as deterministic, but chaotic \cite{DZ_PRE,Moon}. In many cases, random and chaotic processes are hard to distinguish \cite{Rosso}. In particular, some types of chaotic flows are characterized by phase trajectories that can be described as a fractional Brownian motion with $H=0.74$ \cite{Talcinkaya}. This value is close to the one obtained in the presented experiment, indicating that both direct inclusion of noise presented here and chaotic double-oscillator model proposed in \cite{Moon} are well-suited for description of a real clock.

\section{Conclusions}
The dynamics of grasshopper escapement operating in noisy environment has been studied. By using experimental data and numerical simulations, it has been shown that the distribution of pendulum periods is accurately modelled by assuming that the random forces acting on it follow the Maxwell-Boltzmann distribution. The possible origin of this type of statistic is discussed and tested numerically. Moreover, a linear dependence of the clock accuracy on the entropy increase per period is demonstrated, indicating that the recent results regarding quantum and semi-classical nanoscale systems can be extended to classical, macroscopic mechanisms. Finally, the fractal dimension of the function of clock error is calculated, showing its connection with fractional Brownian motion and confirming that the pendulum acts as a system with long-time memory.  

\section*{Appendix A}
\begin{figure}[ht!]
a)\includegraphics[width=.7\linewidth]{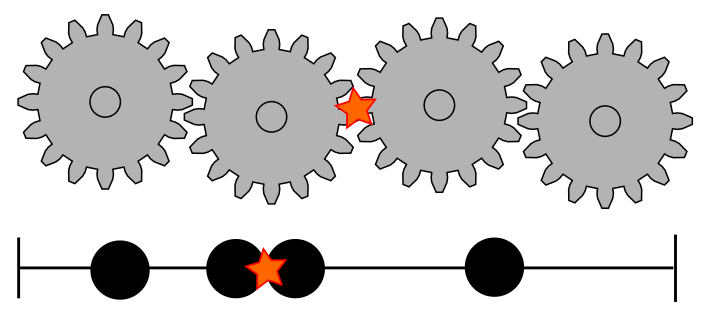}
b)\includegraphics[width=.7\linewidth]{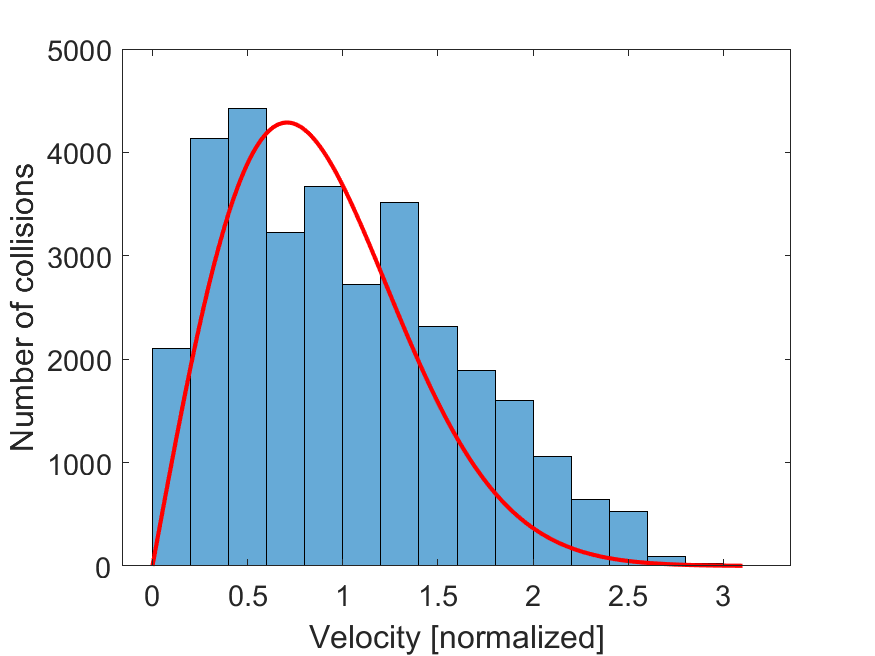}
\caption{a) Schematic representation of the system and its 1-dimensional model. b) Velocity distribution of the particle hitting the wall.}\label{fig_app}
\end{figure}

Lets consider a set of $N$ linked gears, as shown on the Fig. \ref{fig_app} a). 
They are characterized by some value of angular velocity and exchange the angular momentum during collisions. Due to the fact that the angular velocity is a scalar, the motion is one-dimensional and can be modelled as a set of particles moving in one axis (Fig. \ref{fig_app} a), bottom). The system is bound by walls (vertical lines). Due to the friction, the collisions are inelastic, with restitution coefficient $\epsilon=0.999$, e.g. the relative velocity after collision is $0.1\%$ lower than the velocity before the collision. The initial positions of particles $x_i$ are random with uniform distribution $x_0 \in (0,1)$. The walls are located at $x=0$ and $x=1$. Particle diameter is set to $d=0.01$ and they have equal masses. Between collisions, the particles are free and thus follow the equation of motion
\begin{equation}
m\ddot{x}=0.
\end{equation}
After a collision (when distance between two particles is smaller than $d$), for the initial velocities $u_1$, $u_2$, the final velocities $v_1$, $v_2$ are given by
\begin{eqnarray}
v_1 = 0.5(1-\epsilon)u_1 + 0.5(1+\epsilon)u_2,\nonumber\\
v_2 = 0.5(1+\epsilon)u_1 + 0.5(1-\epsilon)u_2.
\end{eqnarray}
Based on the above rules, the motion is numerically integrated with a finite time step $\Delta t=0.01$. To introduce energy to the system, a white noise is added; it has the form of random forces with Gaussian distribution that act on all bodies. Specifically, at every time step a normally distributed value with a mean of 0 and standard deviation of 0.01 is added to every velocity. After some time, an equilibrium is reached where the power of the noise is equal to the power lost in collisions. Let's suppose that one of the walls, e.g. one of the ends of the gear chain, is the escapement. One can calculate the velocity distribution of the particles hitting the wall, and thus the force acting on it. The results calculated for $N=10$ particles are shown on the Fig. \ref{fig_app} b). Despite the relatively small number of particles, the obtained histogram is a fairly close approximation of the Maxwell-Boltzmann distribution (red line). It should be noted that the relatively large number of collisions in the simulation (approximately $3 \cdot 10^4$) is realistic; in real system the contact between gear teeth is not a single continuous force, but series of micro-impacts depending on the surface roughness and high-frequency vibrations of the gears. Furthermore, it has been confirmed that the same distribution is obtained regardless if the noise has normal or uniform distribution.

\end{document}